\documentstyle[aps,twocolumn]{revtex}
\begin{document}
\title{Various Newtonian stresses from macroscopic fluid to microscopic celled fluid}
\author{Gang Liu}
\address{{\small High Performance Computing Virtual Laboratory}\\
{\small Queen's University, Kingston, ON K7L 3N6, Canada}}
\date{November 25, 2004}
\maketitle

\begin{abstract}
We showed that various expressions of stress in different models of fluids
and in different forms of applying Newton's Second Law can be chosen,
however there is always one Newton's Second Law which restricts and provides
flexibility in these expressions. When fluids are regarded as being made of
cells from a microscopic point of view, temperature force (gradient of
internal kinetic energy) can also drive them flow.
\end{abstract}


While the pressure is the normal force acting on a surface per unit area,
the stress means the total force per unit area essentially. However, for a
given system of particle $i$ from $1$ to $n$ with volume $\Omega $, the
stress has been proven or widely used as 
\begin{equation}
\stackrel{\rightharpoonup \rightharpoonup }{\pi }=\stackrel{\rightharpoonup
\rightharpoonup }{\pi _k}+\stackrel{\rightharpoonup \rightharpoonup }{\pi _f}%
,  \label{s01}
\end{equation}
where the kinetic energy term is 
\begin{equation}
\stackrel{\rightharpoonup \rightharpoonup }{\pi _k}=-\frac 1\Omega
\sum_{i=1}^nm_i\dot {{\bf r}}_i\dot {{\bf r}}_i,  \label{s02}
\end{equation}
and the regular force term is 
\begin{equation}
\stackrel{\rightharpoonup \rightharpoonup }{\pi _f}=\frac 1\Omega
\sum_{i>j}\left( {\bf f}_{j\rightarrow i}\right) \left( {\bf r}_j-{\bf r}%
_i\right) ,  \label{s03}
\end{equation}
with mass $m_i$ and position vector ${\bf r}_i$ for particle $i$, and force $%
{\bf f}_{j\rightarrow i}$ acting on particle $i$ by particle $j$. As the
kinetic energy term seems not representing any forces directly, Zhou
recently studied previous derivations for it and concluded that the kinetic
energy term was not physically grounded and should be deleted from the
stress expression\cite{zhou}. In our recent work\cite{lg.eprint1}, we found
that the kinetic energy term can be explained by collision forces between
particles and walls surrounding the system in a statistical way. As there
has been no doubt about the regular force term in principle and we have
discussed it in detail in our previous work for systems with periodic
boundary conditions\cite{lg.eprint1}, we will mainly discuss the kinetic
energy term in views of macroscopic and microscopic motions based on
Newton's Second Law below.

Let us repeat Newton's Second Law first, 
\begin{equation}
\frac d{dt}\left( m_i\dot {{\bf r}}_i\right) ={\bf F}_i\text{, }%
(i=1,2,\cdots ,\text{ }n)\text{ ,}  \label{s04}
\end{equation}
where ${\bf F}_i$ is the net force acting on particle $i$. Even for this
extremely well known law, we still want to emphasis that it means the rate
of the ``regular-force-induced''change of the moment is equal to the regular
force essentially. For one particle, this is never a problem. For many
particles, if we simply add Eq. (\ref{s04}) for all particles and get 
\begin{equation}
\sum_{i=1}^n\frac d{dt}\left( m_i\dot {{\bf r}}_i\right) =\sum_{i=1}^n{\bf F}%
_i\text{, }  \label{s05}
\end{equation}
no problem either. Now let us consider a very short period of time $\Delta
t=t_2-t_1>0$ (from $t_1$ to $t_2$ ) for a system containing the above $n$
particles but defined with a fixed space boundary. Suppose an extra
particle, particle $i+1$, enters the system during this period even with an
unchanged velocity $\dot {{\bf r}}_{n+1}$, and no particles move out of the
system. If we calculate the momentum of the system at $t_1$ and $t_2$ as $%
{\bf P}_1=\sum_{i=1}^nm_i\dot {{\bf r}}_i\left( t_1\right) $ and ${\bf P}%
_2=m_{n+1}\dot {{\bf r}}_{n+1}+\sum_{i=1}^nm_i\dot {{\bf r}}_i\left(
t_2\right) $ respectively, and get the average rate of the momentum increase
of the system for this period as 
\begin{equation}
\frac{{\bf P}_2-{\bf P}_1}{\Delta t}=m_{n+1}\frac{\dot {{\bf r}}_{n+1}}{%
\Delta t}+\sum_{i=1}^nm_i\frac{\dot {{\bf r}}_i\left( t_2\right) -\dot {{\bf %
r}}_i\left( t_1\right) }{\Delta t}\text{, }  \label{s08}
\end{equation}
we need an additional imagined force $\stackrel{\rightarrow }{\sf f}=m_{n+1}%
\dot {{\bf r}}_{n+1}/\Delta t$ acting on the system in order to have
Newton's Second Law satisfied as 
\begin{equation}
\lim_{\Delta t\rightarrow 0}\frac{{\bf P}_2-{\bf P}_1}{\Delta t}=\stackrel{%
\rightarrow }{\sf f}+\sum_{i=1}^n{\bf F}_i\text{, }  \label{s08.01}
\end{equation}
which is equivalent to Eq. (\ref{s05}). The force $\stackrel{\rightarrow }%
{\sf f}$ can be imagined as the interaction between the spaces separated by
the boundary area penetrated by particle $n+1$, as discussed in the
TRANSPORT OF MOMENTUM section in our previous work\cite{lg.eprint1}. So we
have two choices. One is to keep that the rate of moment change in Newton's
Second Law is the rate of the ``regular-force-induced''change of the moment.
Then we have Eq. (\ref{s05}) as Newton's Second Law with the imagined force $%
\stackrel{\rightarrow }{\sf f}$ never needed. The other is that the momentum
change due to any reasons, including particles' passing system boundary, is
accumulated, then we have Eq. (\ref{s08.01}) as Newton's Second Law, and the
imagined forces $\stackrel{\rightarrow }{\sf f}$ must be considered. These
physical views for particle systems are also applicable in complicated cases
of continuous medium/fluid, for which forms of stress are usually derived
and extended to more general cases.

In 1950, Irving and Kirkwood derived their formula for stress similar to the
internal stress below by applying statistics onto hydrodynamics\cite{ik}. We
will apply Newton's Second Law in hydrodynamics and limit statistics to
necessary terms only, and will deal with various stress definitions. It will
be convenient for us to start our derivation by repeating some of their
work. Let us consider a continuous fluid with mass density $\rho \left( {\bf %
r};t\right) $ and ``local velocity'' ${\bf u}\left( {\bf r};t\right) $ at
the point ${\bf r}$ and time $t$. We now imagine $\omega $ to be a space
region somewhere in the interior of the fluid with fixed boundary $S$. Since 
$\omega $ is fixed, the rate of change of momentum in it is 
\begin{equation}
{\bf R}_c=\frac \partial {\partial t}\int_\omega \rho \left( {\bf r}%
;t\right) {\bf u}\left( {\bf r};t\right) d\omega =\int_\omega \frac \partial
{\partial t}\left[ \rho {\bf u}\right] d\omega .  \label{s10}
\end{equation}
The rate of flow of momentum out through the surface of $\omega $ is 
\begin{equation}
{\bf R}_f=\oint_S\rho \left( {\bf r};t\right) {\bf u}\left( {\bf r};t\right) 
{\bf u}\left( {\bf r};t\right) \cdot d{\bf S=}\int_\omega \nabla _{{\bf r}%
}\cdot \left[ \rho {\bf uu}\right] d\omega ,  \label{s11}
\end{equation}
where the direction of the surface $d{\bf S}$ is from inside to outside, and
the surface integral has been converted to a volume integral by Gauss'
theorem. The body force acting on the fluid within $\omega $ due to external
source is 
\begin{equation}
{\bf F}_e=\int_\omega {\bf X}\left( {\bf r};t\right) d\omega ,  \label{s12}
\end{equation}
where ${\bf X}$ is the force per unit volume. The surface force acting on
the fluid within $\omega $ due to neighbor fluid is 
\begin{equation}
{\bf F}_s=\oint_S\stackrel{\rightharpoonup \rightharpoonup }{\pi }\left( 
{\bf r};t\right) \cdot d{\bf S}=\int_\omega \nabla _{{\bf r}}\cdot \stackrel{%
\rightharpoonup \rightharpoonup }{\pi }d\omega .  \label{s13}
\end{equation}
Then as in all hydrodynamical literatures, we have 
\begin{equation}
{\bf R}_c+{\bf R}_f={\bf F}_e+{\bf F}_s,  \label{s14}
\end{equation}
which further results in its differential equivalence of 
\begin{equation}
\frac \partial {\partial t}\left[ \rho {\bf u}\right] +\nabla _{{\bf r}%
}\cdot \left[ \rho {\bf uu}\right] ={\bf X}\left( {\bf r};t\right) +\nabla _{%
{\bf r}}\cdot \stackrel{\rightharpoonup \rightharpoonup }{\pi },  \label{s15}
\end{equation}
since $\omega $ is arbitrary. The last two equivalent equations are usually
called hydrodynamical equation of motion. Actually they are Newton's Second
Law for fluids. Later we will read them in some views and thus get various
definitions of the stress accordingly. But before this, we have to separate
macroscopic motion from microscopic motion.

It is known that each ``smallest'' element of any fluid or material is a
particle and the true motion of it is microscopic motion. Let us use ${\bf v}%
\left( {\bf r};t\right) $ to denote the microscopic velocity of the element.
However we can see their macroscopic motion only. One example is a flowing
river, in which we can see the water is transporting, and the transporting
velocity at each macroscopic point can be measured easily, while the motion
of each individual water molecule is not visible and its microscopic
velocity can not be easily measured. In most cases, the macroscopic and
microscopic velocities are not the same. We can also say the macroscopic
motion is mass transport, and the macroscopic velocity ${\bf u}\left( {\bf p}%
;t\right) $ at a macroscopic point ${\bf p}$ is the velocity of mass of
centre of the portion of the fluid/material in a local small space region $%
\Delta \omega _{{\bf p}}$ around the point. Specifically we have 
\begin{equation}
{\bf u}\left( {\bf p};t\right) \int_{\Delta \omega _{{\bf p}}}\rho \left( 
{\bf r};t\right) d\omega =\int_{\Delta \omega _{{\bf p}}}\rho \left( {\bf r}%
;t\right) {\bf v}\left( {\bf r};t\right) d\omega ,  \label{s16}
\end{equation}
where we use the integration representing the corresponding summation over
particles. In this sense, a group of microscopic particles can still be
regarded as a continuous medium in a small region. Let us define internal
velocity as 
\begin{equation}
{\bf w}\left( {\bf r};t\right) ={\bf v}\left( {\bf r};t\right) -{\bf u}%
\left( {\bf p};t\right) .  \label{s18}
\end{equation}
Then Eq. (\ref{s16}) leads to 
\begin{equation}
\int_{\Delta \omega _{{\bf p}}}\rho \left( {\bf r};t\right) {\bf w}\left( 
{\bf r};t\right) d\omega =0.  \label{s19}
\end{equation}
Let us call the region $\Delta \omega _{{\bf p}}$ as cell ${\bf p}$ and
assume that the whole fluid is made of these full cells without overlapping.
With Eq. (\ref{s19}), the total kinetic energy of the cell ${\bf p}$ , a
microscopic kinetic energy, can be written as 
\begin{eqnarray}
K_{\Delta \omega _{{\bf p}}} &=&\frac 12\int_{\Delta \omega _{{\bf p}}}\rho
\left( {\bf r};t\right) {\bf v}^2\left( {\bf r};t\right) d\omega   \nonumber
\\
&=&K_{\Delta \omega _{{\bf p}},mac}+K_{\Delta \omega _{{\bf p}},int},
\label{s20}
\end{eqnarray}
where the corresponding macroscopic kinetic energy 
\begin{equation}
K_{\Delta \omega _{{\bf p}},mac}=\frac 12{\bf u}^2\left( {\bf p};t\right)
\int_{\Delta \omega _{{\bf p}}}\rho \left( {\bf r};t\right) d\omega ,
\label{s22}
\end{equation}
and the corresponding internal kinetic energy 
\begin{equation}
K_{\Delta \omega _{{\bf p}},int}=\frac 12\int_{\Delta \omega _{{\bf p}}}\rho
\left( {\bf r};t\right) {\bf w}^2\left( {\bf r};t\right) d\omega .
\label{s23}
\end{equation}
It is well known that temperature represents internal kinetic energy. As a
matter of fact, both macroscopic and microscopic velocities and kinetic
energies are dependent on the velocity of the reference coordinate system.
If temperature represented microscopic kinetic energy, it would also be
dependent on the reference coordinate system. On the contrary, the internal
velocity and internal kinetic energy defined above are independent on the
reference coordinate system. So is the temperature representing the internal
kinetic energy. We can not accept a temperature dependent on the reference
coordinate system. Internal stress should also be independent on the
reference coordinate system.

Now let us read every term of Eq. (\ref{s14}), but, for simplicity,
considering the macroscopic motion only. The first left term ${\bf R}_c$ is
the total rate of change of momentum of the system, including the part due
to fluid moving across the boundary, which is not regular-force-induced.
Then the second left term ${\bf R}_f$ is used to cancel this part. So the
total left side is the total rate of change of regular-force-induced
momentum. Then the right side must be pure regular forces, no imagined
force. The external force ${\bf F}_e$ is clear and free from problems. The
surface force ${\bf F}_s$ must be regular forces, so the stress $\stackrel{%
\rightharpoonup \rightharpoonup }{\pi }$ should not have the kinetic energy
part $\stackrel{\rightharpoonup \rightharpoonup }{\pi _k}$ , and should have
the pure regular force part $\stackrel{\rightharpoonup \rightharpoonup }{\pi
_f}$ only.

If the term ${\bf R}_f$ is moved from the left side to the right and put
into the term ${\bf F}_s$, the stress becomes 
\begin{equation}
\stackrel{\rightharpoonup \rightharpoonup }{\pi ^{\prime }}=-\rho {\bf uu}+%
\stackrel{\rightharpoonup \rightharpoonup }{\pi _f},  \label{s24}
\end{equation}
and Eq. (\ref{s14}) changes into 
\begin{equation}
{\bf R}_c={\bf F}_e+\int_\omega \nabla _{{\bf r}}\cdot \stackrel{%
\rightharpoonup \rightharpoonup }{\pi ^{\prime }}d\omega .  \label{s25}
\end{equation}
For stress definition of Eq. (\ref{s24}), the term $-\rho {\bf uu}$
represents the imagined forces between space and space corresponding to
fluid's moving across the boundary, Newton's Second Law must be used in the
form of Eq. (\ref{s25}), and the term ${\bf R}_c$ must include the part due
to fluid moving across the boundary. So the definition of stress depends on
how Newton's Second Law being used.

Next let us read Eq. (\ref{s14}) in view of microscopic motion, the complete
motion. Then, Eq. (\ref{s14}) becomes 
\begin{equation}
{\bf R}_c^{\prime }+{\bf R}_f^{\prime }={\bf F}_e+{\bf F}_s,  \label{s30}
\end{equation}
with ${\bf R}_c^{\prime }=\int_\omega \frac \partial {\partial t}\left[ \rho 
{\bf v}\right] d\omega $ and ${\bf R}_f^{\prime }=\int_\omega \nabla _{{\bf r%
}}\cdot \left[ \rho {\bf vv}\right] d\omega $. The first left term ${\bf R}%
_c^{\prime }$ is the total rate of change of momentum including the part due
to fluid's microscopic moving across the boundary, which is not
regular-force-induced and cancelled by the second left term ${\bf R}%
_f^{\prime }$. Then the right side must be pure regular forces according to
Newton's Second Law, and the stress $\stackrel{\rightharpoonup
\rightharpoonup }{\pi }$ should not have the kinetic energy part. If we move 
${\bf R}_f^{\prime }$ from the left side to the right and put it into the
stress, then the stress can be defined as 
\begin{equation}
\stackrel{\rightharpoonup \rightharpoonup }{\pi ^{\prime \prime }}=-\rho 
{\bf vv}+\stackrel{\rightharpoonup \rightharpoonup }{\pi _f},  \label{s31}
\end{equation}
with the term $-\rho {\bf vv}$ representing the imagined forces between
space and space corresponding to fluid's micro-moving across the boundary,
and Eq. (\ref{s30}) changes into 
\begin{equation}
{\bf R}_c^{\prime }={\bf F}_e+\int_\omega \nabla _{{\bf r}}\cdot \stackrel{%
\rightharpoonup \rightharpoonup }{\pi ^{\prime \prime }}d\omega .
\label{s32}
\end{equation}

Let us now separate the microscopic motion into macroscopic and internal
motions, supposing $\omega $ containing complete cells only. Then we have 
\begin{eqnarray}
{\bf R}_c^{\prime } &=&\int_\omega \frac \partial {\partial t}\left[ \rho 
{\bf v}\right] d\omega   \nonumber \\
\  &=&\int_\omega \frac \partial {\partial t}\left[ \rho {\bf u}\right]
d\omega +\frac \partial {\partial t}\sum_{\Delta \omega _{{\bf p}}\in \omega
}\int_{\Delta \omega _{{\bf p}}}\rho \left( {\bf r};t\right) {\bf w}\left( 
{\bf r};t\right) d\omega   \nonumber \\
\  &=&\int_\omega \frac \partial {\partial t}\left[ \rho {\bf u}\right]
d\omega ={\bf R}_c.  \label{s34}
\end{eqnarray}
Actually the rate of flow of momentum is a statistical idea as early as in
Eq. (\ref{s11}). The first right statistical term $-\rho {\bf vv}$ of Eq. (%
\ref{s31}) for a microscopic particle implies that the particle has the same
probability appearing at any point of a certain space region. Let us assume
the region is the cell in which it resides, then Eq. (\ref{s19}) results in
the averaged internal momentum at each point 
\begin{equation}
\overline{\rho \left( {\bf r};t\right) {\bf w}\left( {\bf r};t\right) }=0,
\label{s35}
\end{equation}
which does not mean no internal motion. Recognizing the average nature in
the rate of microscopic flow of momentum ${\bf R}_f^{\prime }$ and bringing
Eq. (\ref{s35}) into it, we arrive at 
\begin{eqnarray}
{\bf R}_f^{\prime } &=&\oint_S\rho \left( {\bf r};t\right) {\bf v}\left( 
{\bf r};t\right) {\bf v}\left( {\bf r};t\right) \cdot d{\bf S}  \nonumber \\
\  &=&\oint_S\rho \left( {\bf r};t\right) \left[ {\bf u}\left( {\bf r}%
;t\right) +{\bf w}\left( {\bf r};t\right) \right] \left[ {\bf u}\left( {\bf r%
};t\right) +{\bf w}\left( {\bf r};t\right) \right] \cdot d{\bf S}  \nonumber
\\
\  &=&{\bf R}_f+{\bf R}_w,  \label{s36}
\end{eqnarray}
where 
\begin{equation}
{\bf R}_w=\oint_S\rho \left( {\bf r};t\right) {\bf w}\left( {\bf r};t\right) 
{\bf w}\left( {\bf r};t\right) \cdot d{\bf S.}  \label{s37}
\end{equation}
Then Newton's Second Law of Eq. (\ref{s30}) becomes 
\begin{equation}
{\bf R}_c+{\bf R}_f+{\bf R}_w={\bf F}_e+{\bf F}_s.  \label{s38}
\end{equation}
As in the above, one has the choice to move ${\bf R}_f$ from left to right
and define the tress as $\stackrel{\rightharpoonup \rightharpoonup }{\pi
^{\prime }}$ of Eq. (\ref{s24}) or to keep ${\bf R}_f$ in the left. For
either case, we still have a further choice whether to put ${\bf R}_w$ into $%
{\bf F}_s$. If we do so but keep ${\bf R}_f$ in the left, the stress becomes
Irving and Kirkwood's stress\cite{ik} 
\begin{equation}
\stackrel{\rightharpoonup \rightharpoonup }{\pi }_{ik}=-\rho {\bf ww}+%
\stackrel{\rightharpoonup \rightharpoonup }{\pi _f}.  \label{s39}
\end{equation}
This stress will be zero in the example of figure 4 in Zhou's paper\cite
{zhou}, as there is no internal motion in the material. More interesting is
that ${\bf R}_w$ is completely new, when Eq. (\ref{s38}) is compared with
Eq. (\ref{s14}). The right sides of the two equations may be different when
microscopic motion status is considered in Eq. (\ref{s38}). But this
difference is only an accuracy problem if it exists. The term ${\bf R}_w$
arises from different models of the same fluid/material.

Suppose we study macroscopic properties of a fluid or material. As the same
macroscopic motion may come with different internal motions, the macroscopic
properties should be averaged ones over internal motions. For a given cell
with a fixed macroscopic velocity ${\bf u}$, if the direction of the
internal velocity of every microscopic particle is rotated with a fixed
angle, the macroscopic motion is not changed. So let us further assume that
the internal velocity ${\bf w}$ for any given microscopic particle has equal
probability in all directions with the same absolute value. Then for any
given surface ${\bf S}$, we have averaged tensor acting on it 
\begin{equation}
\overline{{\bf ww}}\cdot {\bf S=}\frac 13\left| {\bf w}\right| ^2{\bf S},
\label{s39}
\end{equation}
the averaged internal kinetic energy density 
\begin{equation}
\varepsilon =\frac 12\rho \left| {\bf w}\right| ^2{\bf =}\frac 32\rho 
\overline{{\bf ww}},  \label{s40}
\end{equation}
and finally 
\begin{equation}
\rho \overline{{\bf ww}}=\eta kT,  \label{s41}
\end{equation}
where the relation $\varepsilon =\frac 32\eta kT$ is used with temperature $T
$, Boltzmann constant $k$, and particle number density $\eta $. Please note
that $\rho \overline{{\bf ww}}\cdot {\bf S=}\eta kT{\bf S}$, the total rate
of momentum transferred across the surface ${\bf S}$, is the net rate of
that from inside to outside $\frac 12\eta kT_1{\bf S}$ and that from outside
to inside $\frac 12\eta kT_2{\bf S}$, so $T=\frac 12\left( T_1+T_2\right) $
even if inside temperature $T_1$ and outside temperature $T_2$ are in
different cells and not equal.

Considering this statistical nature, Eq. (\ref{s37}) can be simplified as

\begin{equation}
{\bf R}_w=\oint_S\eta kT\left( {\bf r};t\right) d{\bf S=}\int_\omega k\nabla
_{{\bf p}}\left( \eta T\right) d\omega {\bf ,}  \label{s43}
\end{equation}
which means the net rate of momentum transferred into $\omega $ due to
microscopic particles' internal motion penetrating the boundary. At this
moment it is an imagined force, which can be put inside or outside of stress
definition. Considering the internal motion condition of Eq. (\ref{s19}), we
can easily recognize that the increased momentum due to microscopic
particles' internal moving across the boundary should contribute to the
macroscopic motion in $\omega $, described with ${\bf u}\left( {\bf p}%
;t\right) $.

Now let us consider measurements of the fluid. When some detector is placed
into the fluid to detect/measure some property, the microscopic particles
will inevitably collide with it and the collision must be detected/measured.
If the detector is not moving with the same velocity of the fluid's
macroscopic motion, the collision will be complicated and depend on many
factors. So a point detector moving with the same macroscopic velocity of
the fluid at the point can measure some intrinsic-like properties. In this
case, the collisions between the fluid particles and the detector is due to
particle's internal motion only, and the averaged force in the collisions
can be easily proven as $\rho \overline{{\bf ww}}\cdot \Delta {\bf S=}\eta
kT\Delta {\bf S}$ for the detector's small surface area $\Delta {\bf S}$ 
exposed to the particles, supposing the collisions are complete elastic. So
we found some physical force background for ${\bf R}_w$ or $k\nabla _{{\bf p}%
}\left( \eta T\right) $. If we regard it as an additional regular force, we
can define the internal stress as 
\begin{equation}
\stackrel{\rightharpoonup \rightharpoonup }{\pi }_{int}=-k\eta T\stackrel{%
\rightharpoonup \rightharpoonup }{I}+\stackrel{\rightharpoonup
\rightharpoonup }{\pi _f},  \label{s44}
\end{equation}
with unit tensor $\stackrel{\rightharpoonup \rightharpoonup }{I}$, while
Newton's Second Law becomes 
\begin{equation}
{\bf R}_c+{\bf R}_f={\bf F}_e+{\bf F}_{int},  \label{s45}
\end{equation}
where 
\begin{equation}
{\bf F}_{int}=\oint_S\stackrel{\rightharpoonup \rightharpoonup }{\pi }%
_{int}\left( {\bf r};t\right) \cdot d{\bf S}=\int_\omega \nabla _{{\bf r}%
}\cdot \stackrel{\rightharpoonup \rightharpoonup }{\pi }_{int}d\omega .
\label{s46}
\end{equation}
Essentially $\stackrel{\rightharpoonup \rightharpoonup }{\pi }_{int}$ is the
same as the stress $\stackrel{\rightharpoonup \rightharpoonup }{\pi }$ in
our previous work\cite{lg.eprint1}, where the lattice translation motion is
macroscopic and the motion of particle in the MD cell is internal for every
cell.

As temperature is a fundamental physics quantity, we may also write Newton's
Second Law as 
\begin{equation}
{\bf R}_c+{\bf R}_f={\bf F}_e+{\bf F}_s+{\bf F}_t,  \label{s47}
\end{equation}
where the temperature force ${\bf F}_t=-{\bf R}_w$. A simple experiment can
be done as follows to see the action of this force. Just have a kettle with
half-filled water, boil it a few times to get rid of bubbles, cool it and
let it calm down to no visible motion for a while, and heat it again. Then
the water flow is visible even without bubble and before boiling. This flow
is due to the temperature force, as the temperature is not even.

In summary, we have discussed many possibilities of stress definitions for
different models of the fluid/material and for different forms of using
Newton's Second Law. However any stress definition will be meaningful only
when it is used in Newton's Second Law and Newton's Second Law is used
correctly. By separating macroscopic motion and internal motion from
microscopic motion, we found that the temperature force can also drive a
fluid flow.

{The author wishes to thank Prof. S. S. Wu, Jilin University, Prof.
Ding-Sheng Wang , Institute of Physics, Prof. Si-Yuan Zhang, Changchun
Institute of Applied Chemistry, P.R. China, Dr. Kenneth Edgecombe, Dr.
Hartmut Schmider, Dr. Malcolm J. Stott, and Dr. Kevin Robbie, Queen's
University, Canada, for their helpful discussions and earnest
encouragements. }


\begin{references}
\bibitem{zhou}  M. Zhou, Proc. R. Soc. Lond. A {\bf 459}, 2347 (2003).

\bibitem{lg.eprint1}  E-print, arXiv:cond-mat/0209372.

\bibitem{ik}  J.H. Irving and John G. Kirkwood, J. Chem. Phys. {\bf 18}, 817
(1950).
\end{references}
\end{document}